\documentstyle[prl,aps,epsfig,multicol]{revtex}

\begin{document}

\newcommand{\snn}{\sqrt{s_{NN}}}
\newcommand{\pp}{pp}
\newcommand{\pbarp}{\overline{p}p}
\newcommand{\zvtx}{z_{vtx}}
\newcommand{\spec}{SPEC}
\newcommand{\specp}{{\spec}P}
\newcommand{\specn}{{\spec}N}
\newcommand{\vtx}{VTX}
\newcommand{\vtxt}{{\vtx}T}
\newcommand{\vtxb}{{\vtx}B}

\newcommand{\np}{N_{part}}
\newcommand{\avenp}{\langle\np\rangle}
\newcommand{\npB}{N_{part}^B}
\newcommand{\nc}{N_{coll}}
\newcommand{\avenc}{\langle\nc\rangle}
\newcommand{\half}{\frac{1}{2}}
\newcommand{\halfnp}{(\half\avenp)}

\newcommand{\nch}{N_{ch}}
\newcommand{\etazero}{\eta = 0}
\newcommand{\etaone}{|\eta| < 1}
\newcommand{\dndeta}{d\nch/d\eta}
\newcommand{\dndetazero}{\dndeta|_{\etazero}}
\newcommand{\dndetaone}{\dndeta|_{\etaone}}
\newcommand{\dndetanp}{\dndeta / \halfnp}
\newcommand{\dndetaonp}{\dndeta / \np}
\newcommand{\dndetazeronp}{\dndetazero / \halfnp}
\newcommand{\dndetaonenp}{\dndetaone / \halfnp}
\newcommand{\R} { R_{200/130} }
\newcommand{\Rraw} { R_{200/130}^{raw} }
\newcommand{\nprat} {\np^{200}/\np^{130}}
\newcommand{\Rnp} {\R (\np)}

\title{ Centrality Dependence of 
the Charged Particle Multiplicity
near Mid-Rapidity in Au+Au Collisions at $\snn =$ 130 and 200 GeV }

\author { B.B.Back$^1$, M.D.Baker$^2$, 
D.S.Barton$^2$, R.R.Betts$^6$, R.Bindel$^7$,  
A.Budzanowski$^3$, W.Busza$^4$, A.Carroll$^2$,
J.Corbo$^2$, M.P.Decowski$^4$, 
E.Garcia$^6$, N.George$^1$, K.Gulbrandsen$^4$, 
S.Gushue$^2$, C.Halliwell$^6$, 
J.Hamblen$^8$, C.Henderson$^4$, D.Hicks$^2$, 
D.Hofman$^6$, R.S.Hollis$^6$, R.Ho\l y\'{n}ski$^3$, 
B.Holzman$^2$, A.Iordanova$^6$,
E.Johnson$^8$, J.Kane$^4$, J.Katzy$^{4,6}$, 
N.Khan$^8$, W.Kucewicz$^6$, P.Kulinich$^4$, C.M.Kuo$^5$,
W.T.Lin$^5$, S.Manly$^{8}$,  D.McLeod$^6$, J.Micha\l owski$^3$,
A.Mignerey$^7$, J.M\"ulmenst\"adt$^4$, R.Nouicer$^6$, 
A.Olszewski$^{3}$, R.Pak$^2$, I.C.Park$^8$, 
H.Pernegger$^4$, M.Rafelski$^2$, M.Rbeiz$^4$, C.Reed$^4$, L.P.Remsberg$^2$, 
M.Reuter$^6$, C.Roland$^4$, G.Roland$^4$, L.Rosenberg$^4$, 
J. Sagerer$^6$, P.Sarin$^4$, P.Sawicki$^3$, 
W.Skulski$^8$, 
S.G.Steadman$^4$, P.Steinberg$^2$,
G.S.F.Stephans$^4$,  M.Stodulski$^3$, A.Sukhanov$^2$, 
J.-L.Tang$^5$, R.Teng$^8$, A.Trzupek$^3$, 
C.Vale$^4$, G.J.van Nieuwenhuizen$^4$, 
R.Verdier$^4$, B.Wadsworth$^4$, F.L.H.Wolfs$^8$, B.Wosiek$^3$, 
K.Wo\'{z}niak$^{2,3}$, 
A.H.Wuosmaa$^1$, B.Wys\l ouch$^4$\\
(PHOBOS Collaboration) \\
$^1$ Physics Division, Argonne National Laboratory, Argonne, IL 60439-4843\\
$^2$ Chemistry and C-A Departments, Brookhaven National Laboratory, Upton, NY 11973-5000\\
$^3$ Institute of Nuclear Physics, Krak\'{o}w, Poland\\
$^4$ Laboratory for Nuclear Science, Massachusetts Institute of Technology, Cambridge, MA 02139-4307\\
$^5$ Department of Physics, National Central University, Chung-Li, Taiwan\\
$^6$ Department of Physics, University of Illinois at Chicago, Chicago, IL 60607-7059\\
$^7$ Department of Chemistry, University of Maryland, College Park, MD 20742\\
$^8$ Department of Physics and Astronomy, University of Rochester, Rochester, NY 14627\\
}

\date{\today}
\maketitle

\begin{abstract}

The PHOBOS experiment has measured the charged particle multiplicity
at mid-rapidity in Au+Au collisions at $\snn = 200$ GeV as a function
of the collision centrality. 
Results on $\dndetaone$ divided by the number of participating nucleon pairs
$\avenp/2$ are presented as a function of $\avenp$.
As was found from similar data at $\snn=130$ GeV, the data can
be equally well described by parton saturation models and
two-component fits which include contributions
that scale as $\np$ and the number of binary collisions $\nc$.
We compare the data at the two energies by means of the ratio
$\R$ of the charged particle multiplicity
for the two different energies as a function of $\avenp$.  
For events with $\avenp>100$, 
we find that this ratio is consistent with a
constant value of $1.14\pm0.01(stat.)\pm0.05(syst.)$.

\end{abstract}

\vspace*{-.7cm}

\begin{multicols}{2}
\narrowtext

Collisions of gold nuclei at the Relativistic Heavy-Ion Collider (RHIC) 
at $\snn = 200$ GeV
offer a means to study strongly-interacting matter at high densities and
temperatures.  The goal is to create a large-volume, long-lived state within
which quarks and gluons are no longer confined within hadrons, the
quark-gluon plasma (QGP).
The role of the
collision geometry in determining the initial parton configuration is
important for understanding any collective effects which may be present in
such collisions.
We can study this by means of the mid-rapidity charged particle
multiplicity as a function of the number of nucleons that
participate in the collision, $\np$.  
Measurements of proton-nucleus 
reactions at lower energies\cite{elias} 
suggested that the charged multiplicity from
soft production mechanisms 
should simply scale with $\np$\cite{wounded}.
With increasing energy, however, one might expect some component of particle
production to depend on the number of binary collisions,  
due to the increasing role of hard processes (minijet and jet production).
In nuclear collisions, $\avenc \propto \avenp^{4/3}$, 
making these systems
quite suitable for studying the interplay between the various effects.

Nuclear collisions at RHIC also provide an opportunity to study 
Quantum Chromodynamics (QCD) in a novel
regime where parton densities are high, yet the strong coupling constant
is small due to the large momentum transfers involved.  In such a regime,
gluon densities can be large enough that the gluons recombine, 
causing a saturation of the gluon structure function at low
Bjorken $x$, characterized by a momentum scale $Q_s$. 
Since the parton densities in
the initial state can be related to the density of produced hadrons 
in the final
state, definite predictions are possible regarding the multiplicity of
charged particles as a function of energy and centrality
\cite{kn}.

A recent extension of the calculations by Kharzeev and Levin
\cite{saturation-shapes} has given
predictions of the energy, rapidity, and centrality dependence
of the charged particle multiplicity.
These new calculations use the predicted QCD
evolution of measured results, 
incorporating parameters extracted from 
inclusive deep-inelastic electron-proton scattering data
\cite{golec-biernat}.
Of primary importance in this treatment
is the exponent $\lambda$, which parameterizes
the energy dependence of the saturation scale as
$Q^2_s \propto (\sqrt{s})^\lambda$.
Kharzeev and Levin use this to predict that the energy dependence 
for $\dndeta$ will also scale as ($\sqrt{s})^\lambda$ at high energies.
Furthermore, they predict that the higher energy collisions
allow events with larger impact parameter to be in the saturation regime.  
This affects the multiplicity from peripheral events 
more than for central events, which
already have sufficient parton density at lower energies.
Thus, they predict the multiplicity in peripheral events to rise
slightly faster with energy than for central events.

This effect should be contrasted with the much simpler two-component 
parameterization constructed to interpolate between 
proton-(anti)proton ($\pp,\pbarp$) and central nucleus-nucleus
collisions
by incorporating contributions which scale with the number 
of wounded nucleons
as well as the number of binary collisions.
The pseudorapidity density at mid-rapidity has been
measured  
to rise by 14$\pm$5\% between 130 and 200 GeV
in central $Au+Au$ collisions
\cite{dndeta_200} while
interpolations based on UA5 data on $\pbarp$ collisions at similar
energies \cite{pp} suggest only an 8\% increase for
elementary collisions.  Therefore, the ratio 
$\R = \dndeta|_{200} / \dndeta|_{130}$ should decrease with
increasing impact parameter.

The predictions from the saturation and two-component scenarios 
have been found to
be nearly indistinguishable as a function of centrality at 
$\snn = 130$ 
GeV\cite{kn,saturation-spectra} and agree well with the published
RHIC data from PHENIX\cite{phenix} and PHOBOS\cite{dndeta_cent,dndeta_shapes}.
The authors of Ref. \cite{saturation-spectra} find this to be a nontrivial
consequence of the saturation formalism, although this ambiguity is unlikely 
to be reduced on the basis of the multiplicity data alone.  

We have performed a measurement of the centrality dependence of the 
mid-rapidity charged particle multiplicity in Au+Au collisions at 
$\snn = 200$ GeV using the PHOBOS detector and derived $\R$
as a function of the number of participants.
PHOBOS has previously published results for the centrality 
dependence of $\dndetanp$ at $\snn = 130$ GeV\cite{dndeta_cent} 
as well as $\R$ for the 6\% most central events 
($\avenp \sim 343$)\cite{dndeta_200}. 
The present results are an extension to the previous measurements, 
using similar methods of analysis.

The collision centrality is determined using the signals measured
in two sets of 16 paddle counters (PP and PN) located at $|z|=3.21$ m
with respect to the nominal interaction point.  For
events at $z=0$, these detectors measure
charged particles produced into $3<|\eta|<4.5$.
As discussed in Ref. \cite{dndeta_cent}, we rely on the monotonicity of the
paddle signal with the number of participants (verified by correlations
with the PHOBOS zero-degree calorimeters) to extract $\avenp$ 
for a chosen fraction of the total cross section, based
on the Glauber calculation used by the HIJING model\cite{hijing}. 
The dominant systematic error on this quantity reflects the uncertainty 
on the estimation of the total cross section observed by the PHOBOS
trigger counters.  
By a study of events with a low number of hit paddles using
a full simulation of HIJING events, we have 
estimated the efficiency for triggering on the Au+Au total inelastic cross 
section to be $96\pm3$\% for $\snn = 130$ GeV and $97\pm3$\% for 
$\snn = 200$ GeV.
The slight difference between the two energies stems mainly from 
the increase in the width of the pseudorapidity distribution \cite{dndeta_200}.
In order to reduce the error on the ratio of the multiplicity as a 
function of centrality, we have analyzed the simulations at both
energies in an identical fashion.  This has led to a slightly different
efficiency at $\snn = 130$ GeV than the one presented in Ref. \cite{dndeta_cent}, 
which was $97\pm3\%$.

The charged particle multiplicity has been measured independently for
each energy using the same technique described in \cite{dndeta_cent}, 
based on counting 3-point tracks (``tracklets'') in the spectrometer and 
vertex detectors.  
This approach permits some level of background rejection
relative to simply counting detector hits, as was done in \cite{dndeta_shapes}.
We use a simulation based
on HIJING events and a full GEANT simulation to study the effects of
occupancy, combinatorial background and experimental backgrounds.  
Due to the better granularity in both pseudorapidity
and azimuthal angle in the
spectrometer, its associated systematic error is 4.5\%, to be
compared with 7.5\% in the vertex detector.  As the particle density
does not increase dramatically between the two energies, the systematic
error is the same for the two data sets.  

In forming the ratio $\R$, we have analyzed the data in each
subdetector with the same correction factors for both energies.
This reduces the importance of the precision
of the Monte Carlo simulations used to derive the corrections
at each energy, since they directly cancel in the final value of $\R$.
This is justified by the fact that the particle density does not change
substantially between the two energies.  

We present two forms of $\R$.  The first is done for a chosen fraction
of the total cross section, which requires no other input other than
the trigger efficiency.  The other is the ratio for a constant value of
$\np$, which requires a model of the nuclear geometry.
In both cases, a precise measurement of the $\R$ is achieved by
the average of separate measurements of the ratio in each sub-detector.

\begin{figure}[h]
\begin{center}
\epsfig{file=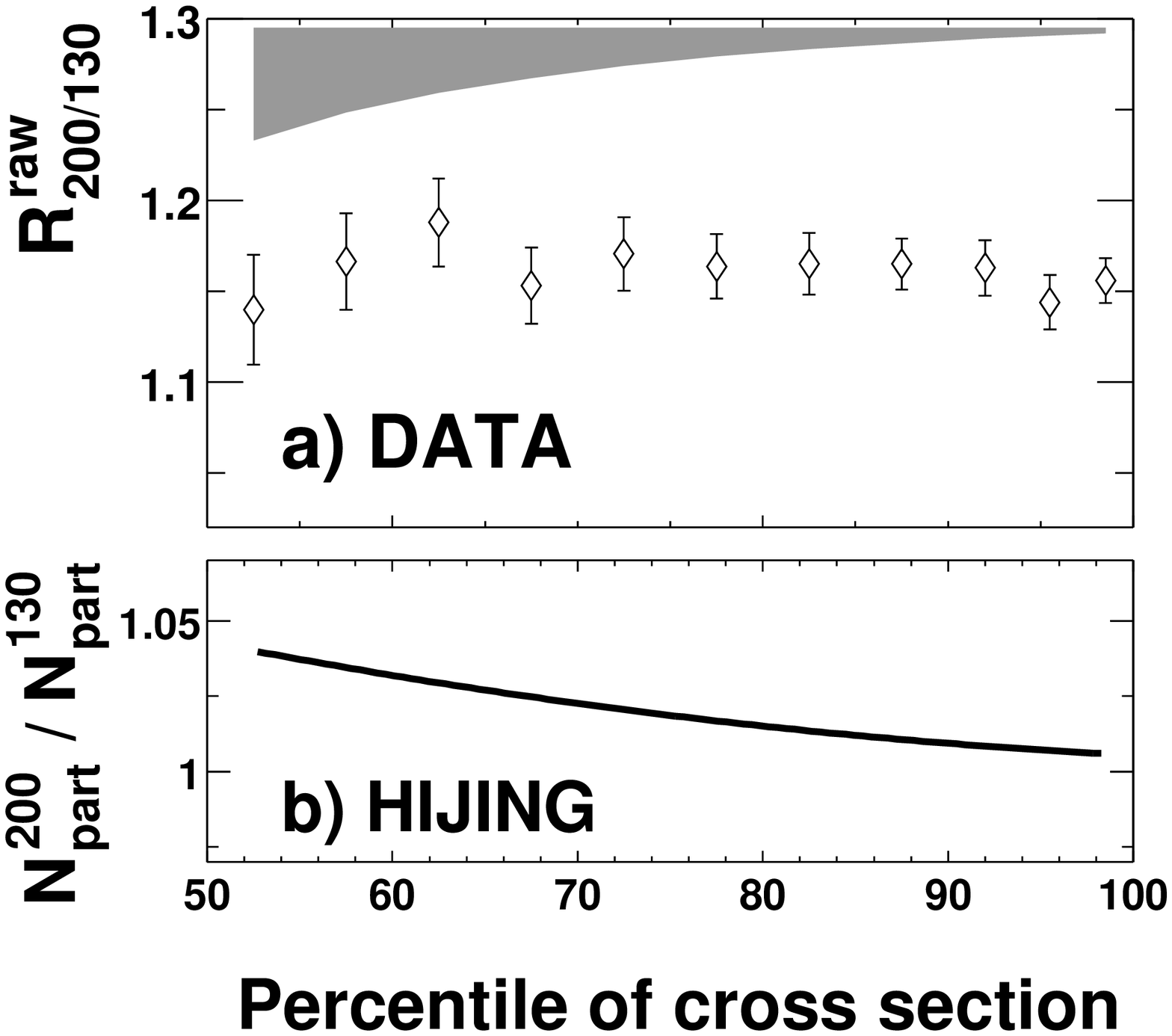,width=9cm}
\end{center}
\caption{
a.) $\Rraw$ vs. the average percentile of cross section.
The grey band indicates the systematic error induced by 
the uncertainty in the relative trigger efficiency between
the two beam energies.
b.) The ratio $\np^{200}/\np^{130}$ for each centrality
bin derived using HIJING at $\snn =$ 130 and 200 GeV.
}
\label{raw_ratio}
\end{figure}

In Fig.\ref{raw_ratio}(a) we show the ratio of $\dndetaone$ at $\snn =
130$ GeV and $200$ GeV for the same percentile of the total cross section,
$\Rraw$.
The values in each bin are listed in the last column of Table 
\ref{results_table}.
This is found to be approximately constant for the 50\% most 
central collisions.
The grey band indicates the magnitude of the systematic error,
which is symmetric around the shown data points.
It reflects the
uncertainty in the estimate of the relative uncertainty of the
trigger efficiency between the two center-of-mass energies.

The ratio of the charged particle multiplicity between $\snn = $ 200 
and 130 GeV as a function of the number of participants
has been obtained simply by
taking the ratio of $\dndetaonenp$ for the two energies,
which corrects for the increase of the inelastic cross 
section between $\snn = 130$ and $200$ GeV (as parameterized in HIJING).
The increasing cross section causes the ratio of $\avenp$ 
between the two energies in each centrality bin, $\nprat$, to increase
by approximately 3\% between the most central events and
the bin corresponding to 45-50\% of the
total cross section ($\avenp=65$), as shown in Fig. \ref{raw_ratio}(b).

It should be emphasized that in both ratio measurements, averaging
over independently-measured ratios does not give precisely the same answer
as that of taking the ratios of the averaged quantity at each beam energy.  
We have chosen the former method since it allows the correction factors
for each measurement to cancel in the ratio.

\begin{figure}[h]
\begin{center}
\epsfig{file=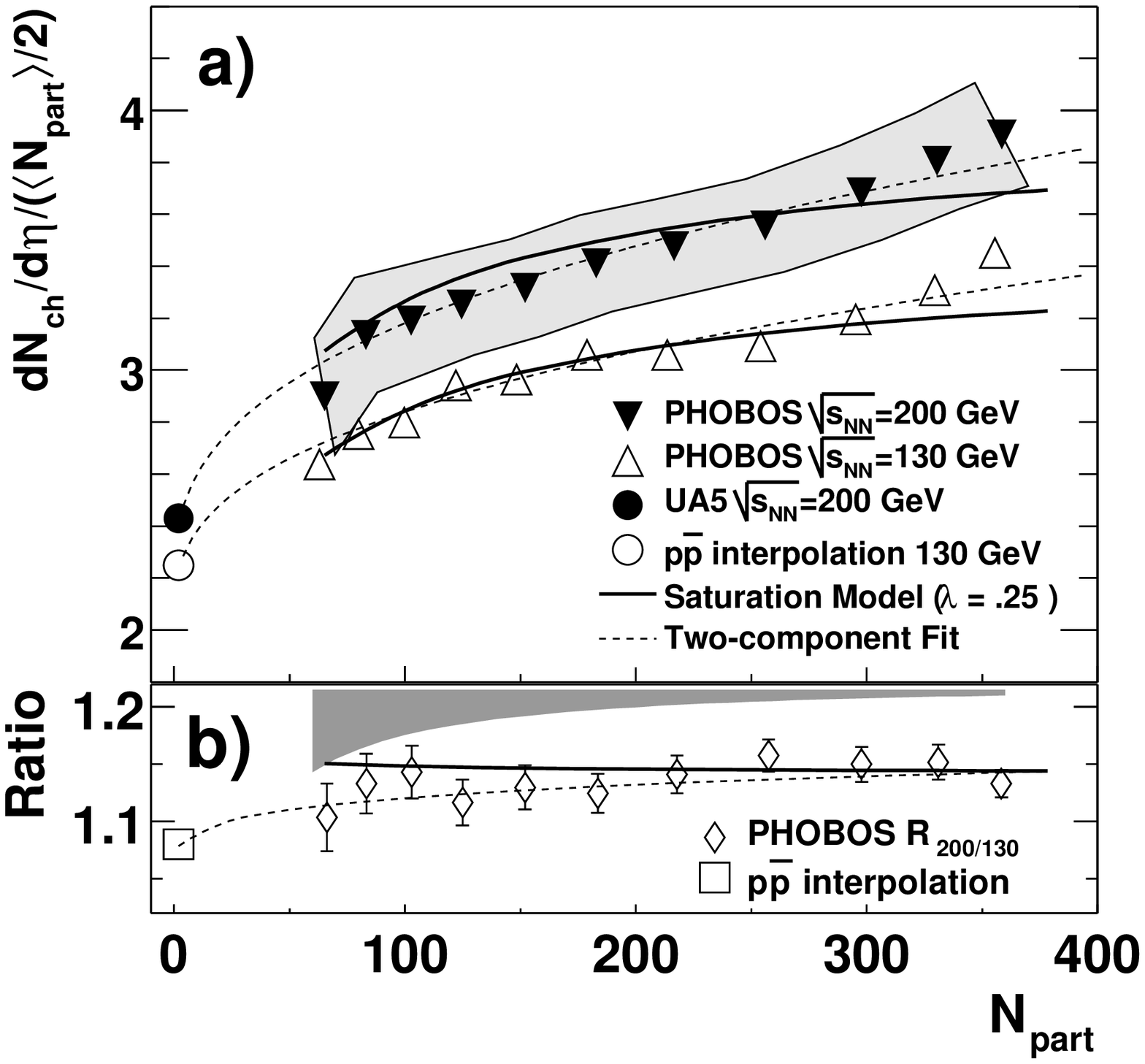,width=9cm}
\end{center}
\caption{
a.) The measured scaled pseudorapidity density $\dndetaonenp$ 
as a function of $\np$ for Au+Au collisions at $\snn =$ 
130 GeV (open triangles) and 200 GeV (closed triangles).
The error band around the 200 GeV data 
combines the error on $\dndetaone$ and $\avenp$.  
The open and solid circles are $\pbarp$ results derived
from the data in Ref. [7].
b.) The ratio of charged multiplicity for $\snn = $ 130 and 200 GeV, $\R$, 
(for constant $\avenp$) vs. $\np$.  
The ratio from $\pbarp$ is shown by the open square.
The grey band indicates the systematic error estimate.
In both panels, results from a saturation model prediction 
and a two-component fit are shown
as solid and dashed lines, respectively.
}
\label{final}
\end{figure}

In Fig. \ref{final}(a) we present the separate results for $\dndetanp$ vs. 
$\np$ for $\snn = 200$ GeV (shown as closed triangles) as well as 
$130$ GeV (open triangles).  
For both energies, we observe a continuous rise of particle density
with increasing centrality.
The total 95\% C.L. error for the 200 GeV data is shown as a shaded band
and includes both statistical and systematic errors.  
The errors at 130 GeV are not shown, but are of similar magnitude.
In Table \ref{results_table}, we list $\avenp$, $\dndetaone$
and $\dndetaonenp$ for each energy and centrality bin.
 
To put these results in context, we also show two calculations.  
The first calculation shown is the prediction of the parton saturation
model for both energies, indicated by solid lines in Fig. \ref{final}(a).  
The only parameter needed to predict the energy
density is the exponent $\lambda$, which is set to $0.25$
as is done in Ref. \cite{saturation-shapes}.  The calculation is truncated at
$\avenp \sim 65$ since this model is not appropriate 
below this value \cite{kn}.

The second calculation involves fits to the data using 
the two-component 
parameterization proposed in Ref. \cite{kn},
$\dndeta = n_{pp}((1-x)\avenp/2 + x\avenc)$.
The parameters have a simple interpretation as
the fraction of production from hard processes ($x$) and the number
of particles associated with a single $pp$ interaction ($n_{pp}$)
when $\np=2$ and $\nc=1$.
We have performed fits using a parameterization of $\avenc$ based on
Ref.\cite{kn}, $\avenc = 0.352 \times \avenp^{1.37}$.
Using the measured values for $n_{pp}$ we find that values of 
$x=0.09\pm0.02$ for $\snn=130$ GeV and
$x=0.11\pm0.02$ for $\snn=200$ GeV account well for the measured 
centrality dependence of $\dndetanp$.  These are shown
as dashed lines in Fig. \ref{final}(a).
The error estimate is based on the allowed range of $x$ which keeps 
the results of the calculation within our stated error bands.
The value of $x$ obtained for $\snn = 130$ GeV agrees with the value
extracted in Ref. \cite{kn}, which only used the $\pbarp$ value
and the central PHOBOS result from Ref. \cite{dndeta_orig}.
The value of $x$ at $\snn = 200$ GeV is consistent with the prediction
in Ref.\cite{saturation-shapes} that $x$ should vary as the square of 
the gluon structure function, $(\snn)^{2\lambda}$, which gives
$x(\snn=200)=1.2 \times x(\snn=130)$ for $\lambda=.25$.
These results are also in good agreement with the two-component calculations
by Li and Wang\cite{shanda}, which use recent parameterizations of
the gluon structure functions as well as nuclear shadowing.

The centrality dependence of $\R$ is shown in Fig. \ref{final}(b) 
and shown in Table \ref{results_table}.
The error bars indicate the contribution from counting statistics, which
becomes important when the systematic effects cancel.  
The systematic error on the centrality dependence of the ratio, 
whose magnitude is shown by the grey band, 
is symmetric around the shown data points 
and incorporates two major contributions.  
The first is the uncertainty in the relative trigger efficiency between the
two energies, which is shown in Fig. \ref{raw_ratio}(a).  
The second error source is the change in $\np$ for a given centrality bin 
as a function of energy, shown in Fig. \ref{raw_ratio}(b).
We have added half of this difference in quadrature
with the first contribution to get the final systematic error.
In addition, there is an overall scale uncertainty on $\R$ and $\Rraw$ of 5\%, 
which was discussed in Ref. \cite{dndeta_200} and stems from the Monte Carlo
studies from which we derived our acceptance, feed-down, and background
corrections.
If we fit the ratio vs. $\np$ to a constant $\R$ above $\avenp=100$, 
we find that it is equal to $1.14 \pm 0.01(stat)$.

For $\np = 2$ we show the ratio of data from $\pbarp$ for $\snn = 200$ GeV 
and an interpolation for $\snn = 130$ GeV\cite{pp}.  We also show the ratios 
of the saturation model predictions (solid line) 
and the two-component fits (dashed line) from 
Fig. \ref{final}(a).  While the two calculations evolve in opposite directions
as the impact parameter increases, they remain sufficiently close 
down to $\avenp \sim 65$ such that our present data cannot 
resolve them definitively.

In conclusion,
the PHOBOS collaboration has measured the pseudorapidity density of charged
particles produced at mid-rapidity in Au+Au collisions at $\snn = 200$
GeV.  
These data have been compared to similar data taken at 
$\snn = 130$ GeV by taking the ratio of multiplicities $\R$ at
a fixed value of $\avenp$.  For $\avenp>100$, we find that this ratio is 
approximately constant at $1.14\pm0.01(stat.)\pm0.05(syst.)$, which is 
consistent both with the parton saturation model predictions and
two-component fits to the data.

This work was partially supported by US DoE grants DE-AC02-98CH10886,
DE-FG02-93ER40802, DE-FC02-94ER40818, DE-FG02-94ER40865, DE-FG02-99ER41099, 
W-31-109-ENG-38 and
NSF grants 9603486, 9722606 and 0072204. 
The Polish groups were partially supported by KBN grant 2 P03B
04916. 
The NCU group was partially supported by NSC of Taiwan under 
contract NSC 89-2112-M-008-024.  We would especially like to thank 
D. Kharzeev and X.-N. Wang for helpful discussions.

\end{multicols}

\newpage

\widetext
\begin{table}
\caption{For each centrality bin, based on the percentile of the total cross section, we show the pseudorapidity density, the number of participants, and the scaled pseudorapidity density for both $\snn = $ 130 and 200 GeV.  The errors shown on these values are systematic.  We also show the ratio $\R$ for each bin both corrected for the different $\avenp$ in each bin, and uncorrected (``raw'').  The errors on the ratios are statistical only.\label{results_table} } 
\begin{tabular}{c|ccc|ccc|cc}
&\multicolumn{3}{c|}{200 GeV}&\multicolumn{3}{c|}{130 GeV}&\multicolumn{2}{c}{Ratios} \\ 

Bin(\%) & $\dndeta$ & $\langle \np \rangle$ & $\dndetanp$ & $\dndeta$ & $\langle \np \rangle$ & $\dndetanp$ & $\R$ & $\Rraw$ \\ \hline 
 0 - 3 & 700 $\pm$ 27 & 358 $\pm$ 12 & 3.91 $\pm$ 0.20 & 613 $\pm$ 24 & 355 $\pm$ 12 & 3.45 $\pm$ 0.17 & 1.13 $\pm$ 0.01 & 1.16 $\pm$ 0.01 \\ 
 3 - 6 & 629 $\pm$ 24 & 331 $\pm$ 10 & 3.81 $\pm$ 0.18 & 545 $\pm$ 21 & 330 $\pm$ 10 & 3.31 $\pm$ 0.16 & 1.15 $\pm$ 0.02 & 1.14 $\pm$ 0.02 \\ 
 6 - 10 & 548 $\pm$ 21 & 298 $\pm$ 9 & 3.68 $\pm$ 0.18 & 472 $\pm$ 18 & 295 $\pm$ 9 & 3.20 $\pm$ 0.16 & 1.15 $\pm$ 0.02 & 1.16 $\pm$ 0.02 \\ 
 10 - 15 & 455 $\pm$ 18 & 256 $\pm$ 8 & 3.56 $\pm$ 0.18 & 393 $\pm$ 15 & 254 $\pm$ 8 & 3.09 $\pm$ 0.16 & 1.16 $\pm$ 0.01 & 1.16 $\pm$ 0.01 \\ 
 15 - 20 & 376 $\pm$ 15 & 217 $\pm$ 8 & 3.48 $\pm$ 0.18 & 327 $\pm$ 13 & 214 $\pm$ 8 & 3.06 $\pm$ 0.16 & 1.14 $\pm$ 0.02 & 1.17 $\pm$ 0.02 \\ 
 20 - 25 & 312 $\pm$ 12 & 183 $\pm$ 7 & 3.41 $\pm$ 0.18 & 274 $\pm$ 11 & 179 $\pm$ 7 & 3.06 $\pm$ 0.17 & 1.12 $\pm$ 0.02 & 1.16 $\pm$ 0.02 \\ 
 25 - 30 & 252 $\pm$ 10 & 152 $\pm$ 6 & 3.32 $\pm$ 0.19 & 220 $\pm$ 8 & 148 $\pm$ 6 & 2.96 $\pm$ 0.17 & 1.13 $\pm$ 0.02 & 1.17 $\pm$ 0.02 \\ 
 30 - 35 & 202 $\pm$ 8 & 124 $\pm$ 6 & 3.25 $\pm$ 0.19 & 180 $\pm$ 7 & 122 $\pm$ 6 & 2.94 $\pm$ 0.18 & 1.12 $\pm$ 0.02 & 1.15 $\pm$ 0.02 \\ 
 35 - 40 & 164 $\pm$ 6 & 103 $\pm$ 5 & 3.19 $\pm$ 0.21 & 140 $\pm$ 5 & 100 $\pm$ 5 & 2.80 $\pm$ 0.18 & 1.14 $\pm$ 0.02 & 1.19 $\pm$ 0.02 \\ 
 40 - 45 & 130 $\pm$ 5 & 83 $\pm$ 5 & 3.14 $\pm$ 0.22 & 110 $\pm$ 4 & 80 $\pm$ 5 & 2.75 $\pm$ 0.20 & 1.13 $\pm$ 0.03 & 1.17 $\pm$ 0.03 \\ 
 45 - 50 & 95 $\pm$ 4 & 65 $\pm$ 4 & 2.90 $\pm$ 0.22 & 83 $\pm$ 3 & 63 $\pm$ 4 & 2.64 $\pm$ 0.21 & 1.10 $\pm$ 0.03 & 1.14 $\pm$ 0.03 \\ 
\end{tabular}

\end{table}
\narrowtext


\begin{references}
\bibitem{elias} J.~E.~Elias, {\it et al.}, Phys.\ Rev.\ D {\bf 22}, 13 (1980).
\bibitem{wounded} A.~Bia\l as, B.~Bleszy\'{n}ski and W.~Czy\.{z}, Nucl.\ Phys.\ {\bf B111} 461 (1976).
\bibitem{kn} D.~Kharzeev and M.~Nardi, Phys.\ Lett.\ B {\bf 507}, 121 (2001).
\bibitem{saturation-shapes} D.~Kharzeev and E.~Levin, Phys.\ Lett.\ B {\bf 523}, 79 (2001).
\bibitem{golec-biernat} K.~Golec-Biernat and M.~Wusthoff, Phys.\ Rev.\ D {\bf 60}, 114023 (1999).
\bibitem{dndeta_200} B.~B.~Back {\it et al.}, Phys.\ Rev.\ Lett.\  {\bf 88}, 022302 (2002).
\bibitem{pp} F.~Abe {\it et al.}, Phys. Rev.\ {\bf D41} 2330 (1990).
\bibitem{saturation-spectra} J.~Schaffner-Bielich, D.~Kharzeev, L.~D.~McLerran and R.~Venugopalan, nucl-th/0108048.
\bibitem{phenix} K.~Adcox {\it et al.}, Phys.\ Rev.\ Lett.\ {\bf 86}, 3500 (2001).
\bibitem{dndeta_cent} B.~B.~Back {\it et al.}, arXiv:nucl-ex/0105011, accepted to Phys.\ Rev.\ C - Rapid Communications.
\bibitem{dndeta_shapes} B.~B.~Back {\it et al.}, Phys.\ Rev.\ Lett.\  {\bf 87}, 102303 (2001).
\bibitem{hijing} X.-N.~Wang and M.~Gyulassy, Phys.\ Rev.\ {\bf D44} 3501 (1991).
\bibitem{dndeta_orig} B.~B.~Back {\it et al.}, Phys.\ Rev.\ Lett.\ {\bf 85}, 3100 (2000).
\bibitem{shanda} S.~y.~Li and X.~N.~Wang, arXiv:nucl-th/0110075.


\end{references}
\end{document}